\documentclass{iau}
\usepackage{graphicx}

\title[Wesenheit Function for Galactic Cepheids] 
{Wesenheit Function for Galactic Cepheids: Application to the Projection Factors}

\author[Ngeow \etal]   
{Chow-Choong Ngeow$^1$, Hilding Neilson$^{2}$, Nicolas Nardetto$^3$ 
 \and Massimo Marengo$^4$}

\affiliation{
$^1$Graduate Institute of Astronomy, National Central University, 
Jhongli City, 32001, Taiwan
\\ {\tt cngeow@astro.ncu.edu.tw} \\[\affilskip]
$^2$Argelander Institute for Astronomy,
Auf dem Huegel 71, 53121 Bonn, Germany\\
$^3$Laboratoire Lagrange,
UMR7293, UNSA/CNRS/OCA, 06300 Nice, France\\
$^4$Department of Physics and Astronomy, Iowa State University,
Ames, IA 50010, USA
}

\pubyear{201{\bf X}}
\volume{289}  
\pagerange{{\bf XXX--XXX}}
\jname{Advancing the physics of cosmic distances}
\editors{R. de Grijs \& G. Bono, eds.}
\begin{document}

\maketitle

\begin{abstract}

Galactic Cepheids are necessary tools for calibrating the period-luminosity relation, but distances to individual Galactic Cepheids are difficult to precisely measure and are limited to a small number of techniques such as direct parallax, main-sequence fitting to open clusters that host Cepheids and Baade-Wesselink (BW) methods. In this work, we re-examine the application of Wesenheit function in determining distances to more than 300 Galactic Cepheids, by taking advantage of the fact that the Wesenheit function is extinction free by definition. The Wesenheit distances were used to calibrate the projection factor ($p$-factor) for Galactic Cepheids that also possess BW distances. Based on $\sim$70 Cepheids, we found that a period-$p$ factor relation may exhibit a non-linear trend with a considerable scatter. During our investigation, discrepant $p$-factors for $\delta$ Cephei were found in the literature. This may be due to inconsistent measurements of the angular diameters using different empirical techniques. We discuss the reason for the inconsistency of angular diameter measurements and offer a possible remedy for this problem. 

Cepheids --- stars: distances --- distance scale
\end{abstract}

\firstsection 

\section{Introduction}

Distances to Galactic classical Cepheids (hereafter Cepheids) have important implication in modern distance scale applications. In contrast to Cepheids in external galaxies that the assumption of equidistance is fulfilled, Cepheids in our Galaxy range in distances from 100 pc to tens of kiloparsecs. Several methods exist in the literature to measure distances to individual Galactic Cepheids. These include: (a) direct parallax measurements (for examples, based on {\it Hipparcos}, {\it HST} and {\it Gaia} in future); (b) Baade-Wesselink (BW) type techniques (which come with several variants, including infrared surface brightness methods, interferometric measurements for angular diameters, CORS and others); (c) main sequence (MS) fitting to open clusters or associations that hosted Cepheids; and (d) light echo technique (currently RS Pup is the only Cepheid with distance measured using this technique). These various methods have only been applied to less than $\sim200$ Cepheids, in some cases that more than one methods are applicable to a given Cepheid. In contrast, more than $\sim1000$ Galactic Cepheids have been recorded to date. In absence of other independent methods, applying a calibrated/theoretical period-luminosity (PL) relation, or a period-luminosity-color (PLC) relation, seems to be the only way to derive distance to Galactic Cepheids. However, applying a PL relation requires the extinction to a given Cepheid is known {\it a priori}, and the error budget in derived distance will have to include the intrinsic dispersion of PL relation (which can be in the order of $\sim0.2$~mag. in optical). Furthermore, the metallicity dependency of PL relation is still under debate. An alternative is to use the Wesenheit function to derive distances to Galactic Cepheids (Opolski 1983; Ngeow 2012), when other independent methods cannot be applied.

\section{The Wesenheit Distance \& the Calibration of Projection Factors}

The Wesenheit function adopted here is in the form of $W=I-1.55(V-I)$. In addition to being extinction free by definition (Madore 1982), this form of Wesenheit function also has the following advantages: (a) its intrinsic dispersion is reduced by $\sim2\times$ to $\sim3\times$ as compared to optical PL relations (Madore \& Freedman 2009; Ngeow \etal~2009); (b) it is linear (Ngeow \etal~2009); and (c) it is insensitive to metallicity (Bono \etal~2010; Majaess \etal~2011). The Wesenheit function used in this work is derived from using $\sim1500$ Cepheids in Large Magellanic Cloud (LMC), based on the superb observations from OGLE-III project, and the intercept has been calibrated with 10 Galactic Cepheids that possess accurate {\it HST} parallaxes (see Ngeow 2012 for more details). Hence, the Wesenheit distance to a given Galactic Cepheid can be calculated using $\mu_W=I-1.55 (V-I) + 3.313 \log(P) + 2.639$, at which the period and mean $VI$ band magnitudes are the only information need to be known. Ngeow (2012) compared the Wesenheit distances for Cepheids that also possess other independent distance measurement (including {\it Hipparcos} parallaxes, MS fitting distances and BW-type distances), with mean differences in distance moduli ranging from $-0.06$~mag. to $0.01$~mag. These results suggested that the Wesenheit distance can indeed be used to derive distance to individual Galactic Cepheids. A large sample of Galactic Cepheids with derived Wesenheit distances can be used to study the metallicity gradient and kinematics of our Galaxy, as well as deriving (multi-band) Galactic PL relations. Some of these applications can be found in Ngeow (2012), and will not be repeated here. In this work, we present the application of Wesenheit distance to the calibration of projection factors (the $p$-factor).

\begin{figure}
\begin{center}
 \includegraphics[width=3.4in]{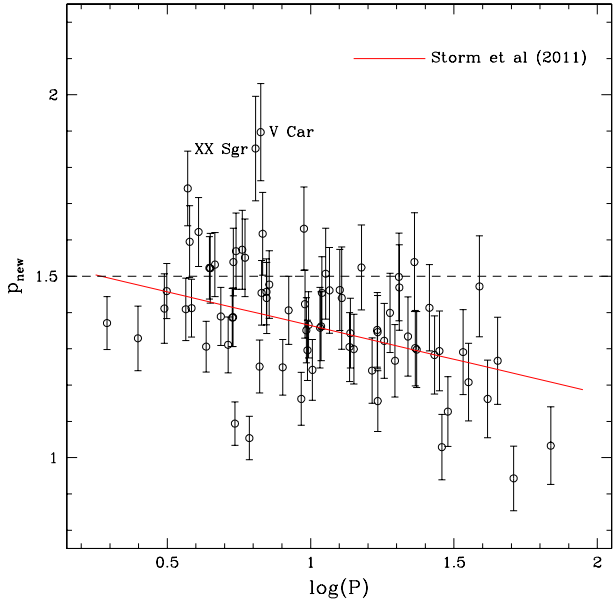} 
 \caption{Calibrated $p$-factors for Galactic Cepheids based on Storm \etal~(2011) sample. The error bars include the errors in both the BW distance and Wesenheit distance. The horizontal dashed line represents the theoretical limit of $p$-factor (Nardetto \etal~2006; Storm \etal~2011). The $Pp$ relation from Storm \etal~(2011) is shown for comparison. A similar plot is also shown in Ngeow \etal~(2012), who calibrated the $p$-factors using a smaller sample of Cepheids, with independent distances from either {\it Hipparcos} parallaxes or MS fitting to the open clusters.}
   \label{fig1}
\end{center}
\end{figure}

The $p$-factor converts the (observed) radial velocity to pulsational velocity, and it is an important parameter in BW-type analysis and/or distance scale application. Since

\begin{eqnarray}
\theta (t) = \theta_0  - \frac{2p}{D} \int [V_r(t)-\gamma ]dt, \label{eqn}
\end{eqnarray}

\noindent the $p$-factor is degenerate with distance $D$ for the same set of observables (the angular diameters $\theta$, radial velocities $V_r$ and gamma velocity $\gamma$). Then, the $p$-factor can be calibrated if a given Cepheid has both BW based distance and an independent distance, i.e. $p_{\mathrm{new}} = p_{BW} \times (D_{\mathrm{indp.}}/D_{BW})$. Figure \ref{fig1} shows the calibrated $p$-factors for a sample of $\sim70$ Galactic Cepheids, where $D_{BW}$ are adopted from Storm \etal~(2011), and $D_{\mathrm{indp.}}$ calculated from Wesenheit distance mentioned previously. This Figure reveals that the period-$p$ factor ($Pp$) relation may not be linear, and may exhibit an intrinsic scatter. Note however that the $p$-factor relation presented in Fig. \ref{fig1} includes also a quantity which is not related to the physics of the p-factor: it includes also by construction the individual discrepancies in our distance indicators (BW and Wesenheit function) which might come also from the intrinsic dispersion of the PL relation or any other bias in the implementation of the methods.

When calibrating the $p$-factors, the $p$-factor for $\delta$ Cephei caught our attention. The $p$-factor given in Storm \etal~(2011), or calibrated here, does not agree with the empirical determination from M{\'e}rand \etal~(2005). Ngeow \etal~(2012) investigated this problem further and found that the derived angular diameters using the infrared surface brightness (IRSB) method from Storm \etal~(2011) and the angular diameters that empirically determined from interferometric technique (M{\'e}rand \etal~2005) do not agree (see Figure 4 in Ngeow \etal~2012). Since the angular diameters are proportional to $p$-factor as shown in equation \ref{eqn} (at the same distance $D$ and using the same radial velocity curve), disagreement of angular diameters naturally lead to the disagreement of $p$-factors. Ngeow \etal~(2012) postulate two possibilities of explaining the disagreement of angular diameters: 

\begin{enumerate}
\item $K$-band flux excess in IRSB Method. This flux excess is presumably due to the existence of circumstellar envelop around $\delta$ Cephei, which could cause the angular diameters to be overestimated by $\sim1\%$.
\item Limb-darkening correction in interferometric technique. Limb-darkening (LD) corrections, derived from plane parallel atmospheres, need to be applied to interferometric measurements. Neilson \etal~(2012) showed that the plane parallel version of the LD corrections can underestimate the angular diameter by $\sim2\%$ when a more appropriate LD correction based on spherically symmetric model atmospheres should be used.  
\end{enumerate}

\begin{figure}
\begin{center}
 \includegraphics[width=3.4in]{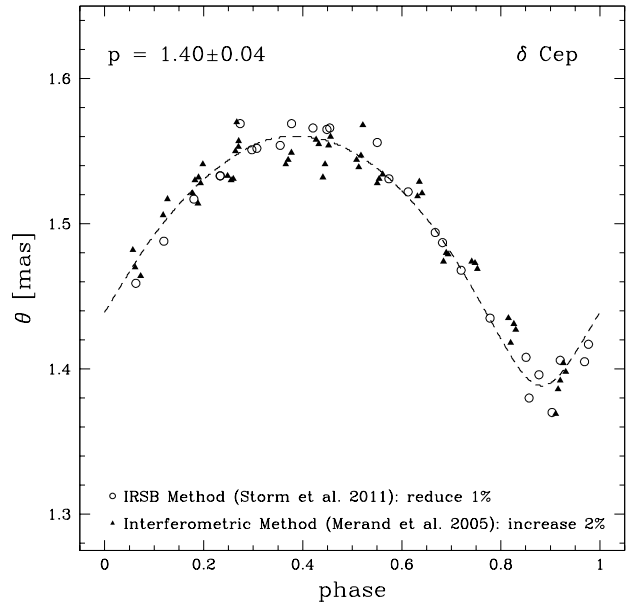} 
 \caption{Comparison of the angular diameters from IRSB method (after reduced the original values by 1\%) and from interferometric measurements (after increased the original values by 2\%). The curves are constructed using equation \ref{eqn} by fitting the angular diameters and $p$-factor (for more details, see Ngeow \etal~2012). The fitted $p$-factor is given in upper-left corner, which is still not consistent with the value by M{\'e}rand \etal\ (2005) of $p=1.27$.}
   \label{fig2}
\end{center}
\end{figure}

To account for these ``biases'', angular diameters from IRSB were reduced by $1\%$ and those from interferometric measurements were increased by $2\%$. The adjusted angular diameters are compared in Figure \ref{fig2}, showing a good agreement after such adjustment. Using equation \ref{eqn} (by adopting the distance D based on {\it HST} parallax), the combined angular diameters can be used to derive the $p$-factor for $\delta$ Cepheid, which is $1.40\pm0.04$. 

\section{Conclusion}

In absence of other independent methods and/or measurements, it is possible to derive the distance to individual Galactic Cepheids using the calibrated Wesenheit function. The derived Wesenheit distances are in good agreement with distance based on other methods (such as {\it Hipparcos} parallaxes, BW based distances and distances from MS fitting), and can be verified using {\it Gaia's} parallaxes in near future. An application of the Wesenheit distance is to calibrate the $p$-factors for Cepheids that also possess BW distances. The calibrated $p$-factors suggested that the $Pp$ relation could be non-linear and may exhibit an intrinsic scatter. For $\delta$ Cephei, the discrepant $p$-factors found in literature, due to the disagreement of angular diameters based on IRSB and interferometric methods, can be remedied if the ``bias'' in both methods can be corrected.

\end{document}